\documentstyle[aps,prd]{revtex}
\begin{document}
\draft
\title{\bf Entropy bounds and black hole remnants\footnote[1]{Based on a
talk at the Conference on Quantum Aspects of Black Holes, ITP, Santa Barbara,
June 1993.}}
\author{Jacob D. Bekenstein\footnote[2]{
Electronic mail: bekenste@vms.huji.ac.il}}
\address{\it Department of Physics, University of California at Santa Barbara,
Santa Barbara, CA 93106\\
and\\
The Racah Institute of Physics, Hebrew University of Jerusalem,
Givat Ram, Jerusalem 91904, Israel\footnote[3]{
Permanent address.}}
\date{Received }
\maketitle
\begin{abstract}

We rederive the universal bound on entropy with the help of black holes  while
allowing for Unruh--Wald buoyancy.  We consider a box full of entropy lowered
towards and then dropped into a  Reissner--Nordstr\"om
black hole in equilibrium with thermal radiation.  We avoid the
approximation that the buoyant pressure varies slowly across the box, and
compute the buoyant force exactly.  We find, in agreement with independent
investigations, that the neutral point generically lies very near the horizon.
A consequence is that in the generic case,
the Unruh--Wald entropy restriction is
neither necessary nor sufficient for enforcement of
the generalized second law.  Another consequence is that generically
the buoyancy makes only a negligible contribution to the energy
bookeeping,  so that the original entropy bound is recovered if the
generalized second law is assumed to hold. The number of
particle species does not figure in the entropy
bound, a point that has caused some
perplexity.  We demonstrate by explicit calculation  that, for arbitrarily
large number of particle species, the bound is indeed satisfied by cavity
thermal radiation in the thermodynamic regime, provided vacuum energies are
included.  We also show directly that thermal radiation in a cavity in $D$
dimensional space also respects the bound regardless of the value of $D$.  As
an
application of the bound we show that it strongly restricts the
information capacity of the posited black hole remnants, so that they cannot
serve to resolve the information paradox.

\end{abstract}

\section{INTRODUCTION}

The generalized second law (GSL) of thermodynamics for black holes
\cite{GSL,GSL2} states that when entropy flows into a black hole, the
sum of black hole entropy and ordinary entropy outside the hole does not
decrease. Arguing from the GSL,  Bekenstein \cite{Bound} has proposed  the
existence of  a universal bound on the
entropy $S$ of any object of maximal radius $R$ and total energy $E$:
\begin{equation}
S\leq {2\pi RE\over \hbar c}.  \label{e1}
\end{equation}
This bound was inferred from the requirement that the  GSL be respected when a
box containing entropy is deposited with no radial motion   next to
the horizon of a Schwarzschild black hole (for a Kerr black hole the condition
is more complex), and then allowed to fall in.  The box's
entropy disappears but an increase in black hole entropy occurs. The
second law is respected provided $S$ is bounded as in Eq.~(\ref{e1}).
Other derivations of the bound based on black holes have been given by
Zaslavskii \cite{Zas,Zas2} and by Li and Liu \cite{LiLiu}. Bound (\ref{e1}) can
also be interpreted as a bound on the information capacity of any object
with total energy $E$ and circumscribing radius $R$
\cite{cost,Numer}.

Bound (\ref{e1}) has been checked directly for quantum fields enclosed in
boxes  of various shapes, when the $S$ is interpreted as the logarithm of the
number of field quantum states up to energy $E$ above the ground state
 (for a review
see Ref.~\cite{Review}).
Numerical checks have
been made for free scalar, electromagnetic and massless spinor fields
enclosed in rectangular or spherical boxes \cite{Numer}.  And an
analytic proof of the bound for those same free fields valid for boxes of
arbitrary shape and topology has been provided \cite{Schif}.
A couple of checks exist for self--interacting fields \cite{Review,Guend}.
All the above demonstrations can be supplemented by the observation that
if the box the fields are enclosed in is reckoned as part of the system,
the bound is even more strongly satisfied because $E$ is thus augmented
while $S$ is hardly changed.
Black holes play no part in any of the above considerations.  Therefore, bound
(\ref{e1}) is known to be true independently of black hole physics for a
variety of systems in which gravity is negligible.

By contrast, just a few pieces of evidence exist concerning bound (\ref{e1})'s
validity  for self--gravitating systems \cite{Zas,Jiu,ZurekPage}.   To these we
may add  that black holes themselves comply with bound (\ref{e1}) if $R$ in
the formula is interpreted as $({\cal A}/4\pi)^{1/2}$, where ${\cal A}$ is the
horizon area \cite{Bound}. Thus our confidence in  bound (\ref{e1}) for
gravitating systems rests almost  entirely on the black hole arguments of
Refs.~\cite{Bound,Zas,Zas2,LiLiu}.

In Ref.~\cite{Bound} it was assumed that the energy at infinity added to the
black hole with the box (which determines the increase in black hole
entropy) is that which may be inferred from the redshift factor at the
deposition point. However, Unruh and Wald (UW) \cite{UW1,UW2}
have pointed out that
when the deposition is attempted by lowering the box from far away,
buoyancy in the radiative black hole environs will prevent lowering the
box
down to the horizon (if one does not wish to invest energy by pushing it
in), and the box will ``float'' at some neutral point.
The total  energy at infinity added to the hole after the box has been
dropped from the neutral point is larger than  the redshifted proper energy
of the box. Accordingly,  UW  concluded that the GSL is  respected when the box
is dropped, provided only its entropy is restricted by
\begin{equation}
S\leq V\,s(E/V),   \label{e2}
\end{equation}
where $s(e)$ is the entropy density as function of
energy density $e$ of {\it unconfined\/} thermal radiation.  UW thus concluded
that bound (\ref{e1}) is unecessary for the proper functioning of the GSL.
They further argued that that bound (\ref{e1}) could not possibly remain
correct when the number of particle species in nature is arbitrarily large.

Although it seems reasonable that thermal radiation maximizes entropy as
a function of energy density, the UW entropy restriction (\ref{e2}) can
easily fail for  a system in which surface effects are sizeable, because such a
system cannot be described entirely in terms of extensive or intensive
variables like $V$ or $E/V$, respectively.  In effect, the shape of the system
is also a variable.  For example, in a rectangular box with dimensions
$d\times d\times 0.1 d$, radiation in the energy range
$[\hbar/d,80\hbar/d]$ explicitly
exceeds the UW restriction on entropy \cite{BcUW2}. [A more dramatic
violation occurs for a black hole; Eq.~(\ref{e2}) predicts a bound on
$S$ that rises like $E^{3/2}$ whereas black hole entropy grows like
$E^2$].
In light of this, UW \cite{UW2} conjectured that the entropy restriction
(\ref{e2}) applies only for box and contents together.  In this revised
form the restriction may be generally correct for non--gravitating
systems; however, it did not play
a role in UW's subsequent discussion of the functioning of the GSL  in
an alternative {\it gedanken\/}
experiment (box emptied into black hole and withdrawn open) \cite{UW2},
and its status has remained unclear.  And because UW's
rescue of the GSL by an appeal to buoyancy \cite{UW1}
in the original {\it gedanken\/} experiment \cite{Bound}
relied on the UW entropy restriction, that experiment has continued to be
problematic for the GSL.

In light of the above, we retrace in Sec.~II.A UW's analysis of
the original {\it gedanken\/} experiment in which
a box full of entropy is lowered towards and then dropped
into a black hole from the neutral point. For later convenience we perform the
{\it gedanken\/} experiment with a Reissner--Nordstr\"om (RN) black hole.  We
studiously avoid UW's approximation that the ambient buoyant pressure varies
slowly across the box because it has become clear \cite{BcUW1,Zas2,LiLiu}
that the neutral point lies very near the horizon where this approximation
must fail because of the large gradients.  Our exact treatment shows that the
UW entropy restriction (\ref{e2}) is neither sufficient nor necessary for the
satisfaction of the GSL.

This situation motivates our rederivation of entropy bound (\ref{e1})
which takes full account of UW buoyancy.  Our new analysis closely
parallels UW's original one, much more so than alternative ones proposed
in the wake of UW's paper \cite{BcUW1,BcUW2,Zas2,LiLiu}; we pinpoint the
stage at which the analysis departs from UW's.  In Sec.~II.B we determine the
position of the neutral point, confirming that it lies very near the horizon.
As a result, the buoyancy makes only a small change in the energy bookeeping.
The original entropy bound, Eq.~(\ref{e1}), is recovered in Sec.~II.C.

As mentioned, the independence of bound (\ref{e1}) on the number of
particle species has been regarded as a sign that it must fail when the
number of species is large \cite{UW1}.  In Sec.~III.A we show by explicit
calculation of thermal radiation entropy in the thermodynamic limit that the
bound is respected for a box containing an
arbitrarily large number of massless particle species,
provided care is taken to include the vacuum energy in the energy of the
full system.  We also show, in Sec.~III.B, that
cavity thermal radiation in the thermodynamic regime in $D$ dimensional
space  also respects the bound regardless of the value of $D$.

Finally, in Sec.~IV we show that the bound strongly restricts the information
capacity of the posited black hole remnants, so that they cannot serve to
resolve the information paradox.

Henceforth we use units with $G=c=k_{\rm Boltzmann}=1$, but continue to display
$\hbar$.

\section{DERIVATION OF THE ENTROPY BOUND}

\subsection{Critique of Unruh and Wald's Analysis}

The system of interest is a macroscopic rectangular box of total energy
$E$ which holds entropy $S$.  We label the horizontal crossectional area
of the box $A$ and its height $b$.  The box is to be lowered towards a
black hole with a face facing it in a standard orientation which is
defined by two technically convenient conditions.  First, we require
that $b$ {\it not\/} be very small compared to $A^{1/2}$.  If this turns
out to be the case for the initial orientation, the box is to be rotated
by $90^{o}$ about a horizontal axis so that a longer edge is brought to
vertical orientation, and that edge is to be labeled by $b$.  Further,
we require that the center of mass (CM) of the box lie initially on the
centroid plane (horizontal plane halfway up the box), or below it.  If
this condition is not satisfied, the box is to be turned upside down
(which respects the previous arrangement), and lowered in that
orientation. We denote by $R$ the circumscribing radius of the box.

For convenience we carry out the {\it gedanken\/} experiment with a
RN black hole of mass $M$ and (not necessarily
electric) charge $Q$.
The exterior metric may be written as
\begin{equation}
ds^2=-\chi^2\,dt^2+\chi^{-2}dr^2+r^2\,(d\theta^2+\sin\theta^2\,d\phi^2),
\label{e3} \end{equation}
where $r$ denotes the usual Schwarzschild radial coordinate and
\begin{equation}
\chi^2={(r-r_{+})(r-r_{-})\over r^2},   \label{e4}
\end{equation}
with $r_{\pm}\equiv M\pm \sqrt{M^2-Q^2}$. The event horizon lies at
$r=r_{+}$ and has area ${\cal A}=4\pi r_{+}^2$ and entropy $S_{\rm
bh}={\cal A}/(4\hbar)$.  In what follows we also employ the radial proper
length
measured from the event horizon, $l$,  defined by $dl=\chi^{-1}\,dr$, and use
the notation $\Delta=r_{+}-r_{-}$.
In order that the box may ultimately be dropped into the hole we assume
that
\begin{equation}
A\ll r_{+}^2;\qquad b\ll r_{+}.
\label{n1}
\end{equation}
We shall assume that the box and its contents are ``transparent'' to
the
gauge field of the hole; thus we do not worry about buoyancy due to
stresses of this field.

Following UW we assume the black hole has reached equilibrium with its
own Hawking radiation, the whole system being enclosed in a large
cavity.  The black hole temperature $T_{\rm bh}$ and the local
temperature $T$ are related by
\begin{equation}
T={T_{\rm bh}\over \chi}={\hbar\Delta\over 4\pi
r_{+}^2\chi}.
 \label{e8}
\end{equation}
The redshift factor $\chi$ enters here as in any equilibrium situation
in a gravitational field.

Applying the first law of thermodynamics to a parcel of equilibrium
radiation, and assuming that its proper enrgy density $e$, pressure $P$,
and proper entropy density $s$ are all functions only of $T$, UW derived
the relation
\begin{equation}
e+P-Ts=0,
\label{n2}
\end{equation}
which merely says that the Gibbs free energy of the parcel vanishes, as
befits a collection of photons or a mixture of equal numbers of neutrinos and
antineutrinos.  Differentiation of Eq.~(\ref{n2}), use of $T=de/ds$,
and simplification with help of Eqs.~(\ref{e8}) and (\ref{n2}) gives
\begin{equation}
d(P\chi)=-e\,d\chi,
\label{n3}
\end{equation}
which is equivalent to the condition of hydrostatic equilibrium
\cite{UW1}.

The buoyant force acting on the box, as measured by an observer at
infinity, is the difference of the redshifted local forces acting on the
upper and lower faces \cite{UW1}:
\begin{equation}
f_{\rm buoy}(l)=A[(P\chi)_{l-b/2}-(P\chi)_{l+b/2}],
\label{n4}
\end{equation}
where $l$ is the proper height of the centroid plane of the box above
the horizon.  UW approximated the difference in this equation by the
first term in a Taylor series.  An exact expression follows from
using Eq.~(\ref{n3}) to convert Eq.~(\ref{n4}) to
\begin{equation}
f_{\rm buoy}(l)=A\int_{l-b/2}^{l+b/2}{e{d\chi\over dl'}}dl'.
\label{n5}
\end{equation}

Let us now express the gravitational force acting on the box and
contents.
The 4--acceleration of a point with 4--velocity $u^\alpha$ is defined
as $a^\alpha=u^\alpha{}_{;\beta}u^{\beta}$.  A simple calculation shows
that a point stationary in Schwarzschild coordinates has invariant
acceleration
\begin{equation}
a\equiv(a^{\alpha}a_{\alpha})^{1/2}=\chi',  \label{e5}
\end{equation}
where a prime denotes  a derivative with respect to $r$.
We can thus write the ``gravitational force'' acting on the box as measured
at infinity in the form
\begin{equation}
f_{\rm grav}(l)=-A\int_{l-b/2}^{l+b/2}{\rho{d\chi\over dl'}}dl'.
\label{n6}
\end{equation}
Here $\rho$ denotes the proper energy density of the box and contents,
and $d\chi/dl=a\chi$ is the local acceleration as measured at infinity.
The minus sign reminds us that gravity and buoyancy act in opposite
senses.

Putting together Eqs.~(\ref{n5}) and (\ref{n6}) we write the work
done by the box on the agent lowering it from infinity down to proper
height $l$ above the horizon as
\begin{equation}
W(l)=\int_{\infty}^{l}{(f_{\rm buoy}+f_{\rm grav})}dl'.
\label{n7}
\end{equation}
$W(l)$ is maximized when the box's centroid plane reaches the neutral
point, $l=l_{0}$.  Setting $dW/dl=0$ we obtain the condition determining
$l_{0}$:
\begin{equation}
f_{\rm buoy}(l_0)+f_{\rm grav}(l_0)=A\int_{l_0-b/2}^{l_0+b/2}{(e-\rho)
{d\chi\over dl}}dl=0.
\label{n8}
\end{equation}

Henceforth, we shall adopt the following notation for integrals like
those appearing in Eqs.~(\ref{n5}), (\ref{n6}) and (\ref{n8}):
\begin{equation}
A\int_{l_0-b/2}^{l_0+b/2}{F}dl\Longleftrightarrow\int_V{F}dV,
\label{extra}
\end{equation}
where $dV$ stands for the element of box volume $A\,dl$.
Note that if $e$ and $d\chi/dl$ are nearly constant across the box, the
condition (\ref{n8}) may be {\it approximated\/} by UW's form, {\it
c.f.,\/}  Eq.~(2.2) of Ref.~\cite{UW1},
\begin{equation}
eV=E\equiv\int_V{\rho}dV,
\label{n9}
\end{equation}
where $V=Ab$ is the box's volume.

Since $W(l)$ is maximum at $l=l_0$, the mass increment $\delta M$ of the
black hole is minimal if the box is dropped in from the neutral point.
Evidently $(\delta M)_{\rm min}=E-W(l_0)$.  Using Eq.~(\ref{n7}) we
may reeexpress this as
\begin{equation}
(\delta M)_{\rm min}=\int_V{\rho\chi}dV+\int_V{P\chi}dV.
\label{n10}
\end{equation}
The first integral is just the energy at infinity of the box, with each
parcel properly redshifted.  This integral equals
$E-\int_\infty^{l_0}{f_{\rm grav}}dl$.
The second integral is just
$-\int_\infty^{l_0}{f_{\rm buoy}}dl$ with $f_{\rm buoy}(l)$ in the form
(\ref{n4}).  As a result of the cancellation of the work done by
buoyant forces on top and bottom of the box over the range
$[l_0+b/2,\infty]$, the buoyant contribution to $(\delta M)_{\rm min}$ only
depends on the distribution of $P\chi$ over the height of the box at the
neutral point.

UW's version of Eq.~(\ref{n10}), namely
\begin{equation}
(\delta M)_{\rm min}=(E+PV)\chi,
\label{n11}
\end{equation}
would follow from Eqs.~(\ref{n9}) and (\ref{n10}) if $P$ and
$\chi$ separately varied little across the box.  Because the neutral
point turns out to be so near the horizon (see Sec.~II.B), these quantities
actually vary a lot across the box in the generic case, so that the
approximations leading to Eq.~(\ref{n11}) are questionable.

Instead, let us replace $P$ in Eq.~(\ref{n10}) by means of the
identity (\ref{n2}).  Taking cognizance of Eq.~(\ref{e8}) we get
\begin{equation}
(\delta M)_{\rm min}=\int_V{(\rho-e)\chi}dV+T_{\rm bh}\int_V{s}dV.
\label{n12}
\end{equation}
Were $\chi$ to be nearly constant across the box, the first integral
here would vanish by virtue of Eq.~(\ref{n9}), and we would be left
with UW's expression (2.22) of Ref.\cite{UW1}.  Since taking $\chi$ as
nearly constant is an unwarranted approximation in view of the closeness
of the neutral point to the horizon, let us instead use
Eqs.~(\ref{n12}) and (\ref{e8}) to compute the overall entropy change of the
world, $(\delta S)_{\rm tot}\equiv
(\delta M)_{\rm min}/T_{\rm bh}-S$, when the
box is dropped from the neutral point.  We find
\begin{equation}
(\delta S)_{\rm tot}=T_{\rm bh}^{-1}\int_V{(\rho-e)\chi}dV+\int_V{s}dV-S.
\label{n13}
\end{equation}

In UW's discussion, the first integral in Eq.~({\ref{n13})
dropped out in wake of the indicated approximations.
It may be seen by comparing
with Eq.~(\ref{n8}) that,  in fact, its magnitude and even its sign
depend on the distribution of $e$ and $\rho$ across the box.  Thus,
in contrast with UW's discussion, we must conclude that validity of the
UW entropy restriction, Eq.~(\ref{e2}), does not by itself guarantee
that the GSL will be obeyed.  And conversely, assuming that the GSL is
satisfied in the {\it gedanken\/} experiment in question, does not allow
us to derive the UW entropy restriction for the box.  Unless
supplemented by detailed information about the box, the entropy
restriction is neither a necessary nor a sufficient condition for the
operation of the GSL.  Therefore, it seems best to calculate in detail the
sum of integrals in Eq.~({\ref{n13}).

\subsection{The Neutral Point}

As the first step we make an approximate determination of the location
of the neutral point by using UW's criterion Eq.~(\ref{n9}), and a
model of radiation as a mixture of noninteracting gases of massless
particles, one for each species in nature.  According to Boltzmann, for
such radiation
\begin{equation}
P=e/3={N\pi^2T^4\over 45\hbar^3},
\label{e7}
\end{equation}
where $N$ is the effective number of particle species (photon and
graviton contribute one to $N$, while each neutrino and antineutrino
species counts as $7/16$).

With Eq.~(\ref{e8}) for $T$, criterion (\ref{n9}) amounts to
\begin{equation}
\chi(l_0)=0.0717N^{1/4}(Ab\hbar/E)^{1/4}\Delta r_{+}^{-2}.
\label{n15}
\end{equation}
Since we assume the box to be macroscopic, and hence large compared to its
Compton wavelength, $\hbar/E\ll b$.  This together with inequalities
(\ref{n1}) tells us that for realistic values of $N$ (in Sec.~IIIA we
shall reconsider the situation where $N$ is arbitrarily large),
$\chi(l_0)\ll\Delta r_{+}^{-1}$. A look at Eq.~(\ref{e4}) shows that the
neutral point must lie in the region $r-r_{+}\ll\Delta$.  But in this
region, a  good approximation to the proper distance
from the horizon to a point $r$ is
\begin{equation}
l\approx 2r_{+}(r-r_{+})^{1/2}\Delta^{-1/2}.  \label{e10}
\end{equation}
It is clear that in that region $l\ll r_{+}$.  In the same approximation
Eq.~(\ref{e4}) can be written
\begin{equation}
\chi(l)\approx {1\over 2}\Delta r_{+}^{-2}l.  \label{e11}
\end{equation}

Two important consequences follow from the linear form of $\chi$ in the
near horizon region where the neutral point lies.  First from
Eq.~(\ref{e8})
\begin{equation}
T=\hbar/(2\pi l),
\label{n16}
\end{equation}
so that, as a function of proper distance from the  horizon, $T$ does not
depend on the black hole's parameters. This universal form for $T$ may
be understood by writing down the Unruh temperature for an (accelerated)
sationary observer, $T_{\rm U}=\hbar a/(2\pi)$, and using Eqs.~(\ref{e5})
and (\ref{e11}) to recast it into a form identical to Eq.~(\ref{n16}).
The equality $T=T_{\rm U}$ in the near horizon region allows one to
interpret the thermal radiation in equilibrium with the hole as observer
dependent Unruh acceleration radiation seen by the stationary
observers.  One may thus follow UW in asserting that the large $e$ and $P$
corresponding to $T$ at the neutral point do not generate strong curvature,
{\it i.e.,\/} that the metric (\ref{e3}) remains an excellent approximation.

The second consequence of the $\chi\propto l$ form is that condition
(\ref{n8}) can be rewritten as
\begin{equation}
\int_V{e}dV=E,
\label{n17}
\end{equation}
which, unlike the UW condition (\ref{n9}), is accurate in the near
horizon region, and indeed becomes the more accurate the nearer to the
horizon since Eq.~(\ref{e11}) is asymptotically exact there.

Let us now use Eq.~(\ref{n17}) to accurately determine $l_0$.  Substituting $e$
from Eq.~(\ref{e7}) and $T$ from Eq.~(\ref{n16}), and remembering
that the range of $l$ in the integral is $[l_0-b/2,l_0+b/2]$, the
condition leads to
\begin{equation}
{(l_0^2-b^2/4)^3\over
3l_0^2b^4+b^6/4}=
\eta^3\equiv{N\hbar A\over 720\pi^2 Eb^3}
\label{e12}
\end{equation}
In our world where the number of particle species that would be excited by a
massive black hole is limited, $\eta\ll 1$
because $\hbar/E\ll b$  and because in the standard box orientation, $b^2$
cannot be very small compared to $A$.
In Sec.~III.A we shall argue that $\eta\ll 1$
remains true even when $N$ is arbitrarily large.

When $\eta\ll 1$, it follows from Eq.~(\ref{e12})  to lowest order in
$\eta$ that
\begin{equation}
l_0^2\approx (1/4+\eta)b^2.    \label{e14}
\end{equation}
Therefore, at the neutral point,
the centroid plane of the box is just a little farther from the
horizon than half the box's height, which means the box floats almost touching
the horizon. This phenomenon has been noted earlier \cite{BcUW1,Zas2,LiLiu}.
Since the neutral point does occur near the horizon, the approximations
involved in Eqs.~(\ref{e10}) and (\ref{e11}) are seen to be justified
{\it a posteriori\/}.  The finding here serves to emphasize that the UW
neutral point condition (\ref{n9}) cannot be accurate because
of the strong gradient of $e$ near the horizon, and should properly be
replaced by condition (\ref{n17}).

Of course, all the above hinges on the correctness of the Boltzmann
model, Eq.~(\ref{e7}).  Although the appearance of new species as
$T$ rises is known not to cause a measurable departure from the
Boltzmann ``equation of state'', but merely requires that $N$ be
regarded as slowly rising, at sufficiently high $T$ interactions may
cause a substantial departure from the simple relation (\ref{e7}).
How would this affect our discussion ?

Let us, in the manner of high energy physics, characterize the onset of
a strong departure from the Boltzmann model by a scale of length $L$
which is connected to the appropriate transition temperature by a
relation much as Eq.~(\ref{n16}).  We know that $L<10^{-12}$cm
because Eq.~(\ref{e7}) works well in cosmology all the way back to
the lepton era.  If the strong interaction is the one that spoils the
Boltzmann model, we may expect $L\leq 10^{-13}$cm.  At any rate, it is
clear that $L$ is much smaller than the dimensions of a macroscopic box.

A departure from Boltzmann's model affects our discussion only if at the
neutral point the  lower side of the box has penetrated to within
$l<L$, so that it is exposed to temperatures beyond the transition.  But,
of course, since the box is much larger than $L$, this means
that at the neutral point the box must nearly touch the horizon, with
its centroid plane at proper height $\approx b/2$, just as we
found by relying on the Boltzmann model.  Thus, as far as the location of
the neutral point is concerned, complications beyond the Boltzmann model
are of no practical consequence.

\subsection{The Entropy Bound}

Proceeding with our program, we now compute explicitly $\delta S_{\rm bh}$ by
going back to Eq.~(\ref{n10}).  Due to the linearity of $\chi$ with $l$, the
integral over $\rho$, the energy density in the box, may be expressed in terms
of the proper height of its CM above the horizon, $l_{\rm cm}$, defined by
${1\over 2}\Delta r_{+}^{-2} l_{\rm cm}=\overline\chi|_{l_0}$ (average
with respect to $\rho$), and $E$ as
defined by Eq.~(\ref{n9}).  And the integral over $P$ may be worked out with
help of Boltzmann's model, Eq.~(\ref{e7}), and
Eqs.~(\ref{e10})-
(\ref{n16}).  The result is
\begin{equation}
(\delta M)_{\rm min}={1\over 2}\Delta r_{+}^{-2} E\ [l_{\rm cm} + \eta^3 l_0
b^4(l_0^2-b^2/4)^{-2}].
\label{e16}
\end{equation}
 A look at Eq.~(\ref{e14}) shows that  the square parenthesis in
Eq.~(\ref{e16}) amounts to $l_{\rm cm}+\eta l_0$; the second term comes from
the buoyancy.  We recall now that in the standard box orientation employed
here, the CM of the box cannot lie above the centroid plane
($l_{\rm cm}\leq l_0$ initially, and the CM can only
slip down as the box enters strong gravitational fields).  And because
$l_0\approx b/2+\eta b$ [see Eq.~(\ref{e14})], the square parenthesis in
Eq.~(\ref{e16}) cannot exceed  $(1+3\eta)b/2$.  The buoyancy corrections of
O$(\eta)$ are evidently negligible under the same assumptions that led
us to conclude that the neutral point is near the horizon.  We thus obtain
\begin{equation}
(\delta M)_{\rm min} \leq
{1\over 4}\Delta r_{+}^{-2}b E,
\label{n18}
\end{equation}
which is another version of Eq.~(\ref{n12}).

{}From Eq.~(\ref{n18}) we may  compute,
with help of the first law for black holes, Eq.~(\ref{e8}) and the
obvious constraint $b<2R$ that when the box is dropped
from the neutral point,
\begin{equation}
(\delta S)_{\rm tot} <
2\pi RE/\hbar - S,
\label{n19}
\end{equation}
which replaces Eq.~(\ref{n13}).
It is evident from this inequality
that in order for the GSL to be satisfied [$(\delta S)_{\rm tot}\geq 0]$, the
box entropy $S$ must satisfy the bound on entropy (\ref{e1}).

The above argument requires modification if at the neutral point the box
already penetrates into the region with $T$ beyond the transition
temperature at which the Boltzmann model gets modified. We first note
that  whatever the true relation between $e$, $P$ and
$T$, the inequality $P<e$ must be satisfied.  The reason is that
causality demands that $dP/de\leq1$ (speed of sound in the radiation
subluminal) for all $T$.  Integrating the inequality $dP-de\leq 0$ from
low $T$ where $P=e/3$ certainly applies, we see that $P<e$ for any T,
however large.
Consider now the identity
\begin{equation}
d(P\chi l)=ld(P\chi)+P\chi\,dl.
\label{n20}
\end{equation}
If we substitute in it from Eq.~(\ref{n3}), and take into account that
in the near horizon region $l\,d\chi=\chi\,dl$, we see that $P<e$
implies that $d(P\chi l)/dl<0$ near the horizon.  This means that $P(l)$
drops off with $l$ faster than $l^{-2}$, which means that whatever the
modification to the Boltzmann model, $P(T)$ must grow faster than $T^2$.

In light of the above, consider the buoyant contribution to $(\delta
M)_{\rm min}$ in Eq.~(\ref{n10}), written as
\begin{equation}
\int_V{P\chi}dV={\int_V{P\chi}dV\over\int_V{P}dV}\cdot\int_V{P}dV.
\label{n21}
\end{equation}
By the causality constraint, $\int_V{P}dV<E$ [see Eq.~(\ref{n17})].  The
ratio of integrals is evidently largest when $P$ decreases slowest with
$l$.  We may thus bound from above that ratio by using the limiting form
$P\propto l^{-2}$ to compute it.  Recalling Eq.~(\ref{extra}) we thus
have
\begin{equation}
\int_V{P\chi}dV<{1\over 2}\Delta r_{+}^{-2} E(l_0+b/2){\ln y\over y-1},
\label{n22}
\end{equation}
where $y\equiv (l_0+b/2)(l_0-b/2)^{-1}$.  The closeness of the box's
bottom to the horizon at the floating point means that $l_0\approx b/2$;
therefore, $l_0+b/2\approx 2 l_0$ and $y\gg 1$.  Hence,
\begin{equation}
\int_V{P\chi}dV=\varpi \Delta r_{+}^{-2}l_0 E,
\label{n23}
\end{equation}
where $\varpi < {\ln y/(y-1)}\ll 1$.

 Inserting this result in Eq.~(\ref{n10}) for $(\delta M)_{\rm min}$ and
recalling that the integral over $\rho$ is already evaluated in our
previous result (\ref{e16}), we have
\begin{equation}
(\delta M)_{\rm min}={1\over 2}\Delta r_{+}^{-2}E
\ [l_{\rm cm}+2\varpi l_0].
\label{n24}
\end{equation}
Again, because the box is standardly oriented and very close to the horizon,
$l_{\rm cm} \leq l_0\approx b/2\leq R$.   And because $\varpi\ll 1$, the
buoyant  term evidently cannot make the square brackets larger than
$R$.  We thus find that  Eqs.~(\ref{n18}) and (\ref{n19}) apply again,
and the entropy bound (\ref{e1}) follows from the assumed validity of the GSL,
as in the case when the Boltzmann model could be used down to the
neutral point.

In all arguments in this subsection, a slightly tighter  bound on entropy
would follow if we worked throughout in terms of $b$, rather than
appealing to the inequality  $b<2R$.  The impression we get from this
that the maximal entropy of a thin box decreases with decreasing
thickness $b$ is supported by numerical computations \cite{Numer} if
$b$ is not too small.  Those calculations do not support Li and Liu's
conclusion \cite{LiLiu} that the entropy bound is unconditionally set by
the smallest box dimension.  After all, in 3--dimensional space a
2--dimensional box filled with massless quanta can hold nonvanishing
entropy.  In our approach here the bound could not be derived in terms
of $b$ for arbitrarily small $b$ because the condition $\eta\ll 1$ will
fail when $b$ gets sufficiently small [see Eq.~(\ref{e12})].

\section{EXTENSIONS}

\subsection{Irrelevance of Species Number}

Our derivation of the bound in Sec.~II assumed $\eta\ll 1$, which
condition would
seem to fail
in a world where $N$ is very large.  Indeed, UW contended that no bound
of type (\ref{e1}), which is independent of the number of species, could
possibly retain its validity as $N$ becomes large because it is known
that the more species there are, the larger the number of states
(entropy) accessible with given energy.  Their own
entropy restriction, Eq.~(\ref{e2}),
scales as $N^{1/4}$, that being the dependence of Boltzmann's formula for
blackbody radiation entropy at given energy on the number of species.
UW  conceived of buoyancy,
with its intrinsic dependence on the number of
species in the radiation, as nature's exclusive
way to defend the GSL against a
violation when a body made up of many species (large entropy with
moderate energy) is lost down a black hole.

That this view must have restricted validity is clear from the following
argument demonstrating that an $N$--independent entropy bound must exist
in order
for the GSL to function, even in situations where buoyancy cannot play a
role.

Consider a box with energy $E$ and entropy $S$ dropped freely
{\it from far away\/} into an {\it exactly extremal\/}
RN black hole of mass $M$ in empty space.
Because the fall is free {\it i.e.,\/} geodesic,
the box does not feel Unruh radiation.  And
because the hole is extremal, thermal radiant pressure is  absent.
Hence there is no buoyancy to
complicate the energy bookeeping.
Initially $S_{\rm bh}=\pi M^2/\hbar$.
After
the black hole has assimilated the box, the mass has gone up to $M+E$ and
$S_{\rm bh}= \pi(M+E+\sqrt{2ME+E^2})^2/\hbar$ (because $Q$ is unchanged).  We
regard $E$ as a small quantity.  Then to lowest order in $E$,
\begin{equation}
\delta S_{\rm bh}=\pi (2M)^{3/2} E^{1/2}/\hbar. \label{e19}
\end{equation}

Coherent gravitational and electromagnetic radiation may be emitted due to the
infall.  The energy in these radiations is expected to be of order $E^2/M$
\cite{Ruff}.  Its substraction from the final $M$ will produce terms in
Eq.~(\ref{e19}) of the same order as those we have already discarded.  Thus the
infall radiations prove negligible, and in any case they would act to reduce
$\delta S_{\rm bh}$. There may also be some radiation entropy
emitted by the Hawking process as the box is assimilated and the  black hole
departs from exact extremality.  Once the black hole is not extreme, it
should radiate entropy at a rate a bit higher than a blackbody of
temperature $T_{\rm bh}$ and area $\cal A$ \cite{Page}.  This entropy
emission should be somewhat larger than the induced decrease in $S_{\rm
bh}$.  Therefore, the
overall entropy growth rate would be
\begin{equation}
dS_{\rm bh}/dt+dS_{\rm rad}/dt\sim (180)^{-1}\Delta^3r_{+}^{-4}.
\label{extra2}
\end{equation}
In our case, after the assimilation
$\Delta=2(2ME)^{1/2}$.  Therefore, over a period of coordinate time $\sim
M$, which is of the order of the time required for the effective
disappearance of the box,
the just perturbed black hole should generate entropy $\sim (4\surd
2/45)(E/M)^{3/2}$,
which is certainly much smaller than $\delta S_{\rm bh}$ because $M$ has to
be large compared to the Planck mass.

Thus we have accounted for all entropy
contributions. In order for the GSL not to be violated,  the box's
entropy $S$ must be bounded by  $\pi (2M)^{3/2} E^{1/2}/\hbar$.  Of
course, for large $M$ this is a much larger bound than
Eq.~(\ref{e1}).  However, the point is
that the new bound, though derived without recourse to lowering into the
black hole,  is also independent of $N$.
This shows that there must exist an
$N$--independent bound on the entropy of a bounded system,
arguments involving buoyancy notwithstanding.

This said, we still face a paradox: by Boltzmann's formulae it
seems that for large enough $N$, a quantity of radiation with given
energy should surpass any $N$--independent entropy bound.  The resolution we
develop here depends on recognizing that the bounds on entropy derivable with
the help of black holes must always refer to an entire system, not to  part of
one.  In particular, bound~(\ref{e1}) refers to the entropy and energy of
the box {\it and} its contents. Therefore, when comparing thermal radiation
entropy in a box with the bound, one would like to restate the Boltzmann
formula
for entropy in a form that takes cognizance of the existence of the box and
those of its properties which are responsible for confining the
radiation.  Then a comparison can be made.

We shall perform the concrete calculations for a spherical box
of radius $R$ and total energy $E$.  According to Boltzmann,
at temperature $T$ thermal
radiation in the box has energy
\begin{equation}
E_{\rm rad}=(4\pi^3/45)NR^3T^4\hbar^{-3}  \label{e20}
\end{equation}
and entropy
\begin{equation}
S_{\rm rad}=(16\pi^3/135)NR^3 T^3\hbar^{-3}.  \label{e21}
\end{equation}
If $E_0$ stands for the energy of the empty box ($E$ at $T=0$), then
upon eliminating $T$ we have
\begin{equation}
S=(4\surd 2/135)N^{1/4}[45\pi R(E-E_0)/\hbar]^{3/4}. \label{e22}
\end{equation}

How small can $E_0$ be ?  Even when empty, the box's energy
receives a contribution from vacuum (Casimir) energy of those fields it
can entrap (or keep out).  On dimensional grounds each species
contributes vacuum energy $\varepsilon=\alpha\hbar/R$ where
$\alpha$ is dimensionless.  For the electromagnetic field in a sphere
Boyer \cite{Boyer} showed that $\alpha=0.045$. For all species
together we write $\varepsilon=\,\bar\alpha N\hbar/R$, where $\bar\alpha$ is
the suitable average.  If positive, the vacuum energy sets a lower bound on
$E_0$.

Even when the vacuum energy is negative \cite{Luk,Unwin}, the total box mass
must end up being  positive.  The physical mechanism is the suction on the
box's wall that must accompany negative vacuum energy.  In
fact, from the rate at which $\varepsilon$ gets more
negative as the box's radius decreases, it follows that the wall
sustains a negative pressure $P=-\bar\alpha N\hbar/4\pi R^3$.
To resist this suction the wall must maintain a surface pressure (force
per unit length) of  the same order \cite{BcUW1}.
But unless the wall's surface mass
density is bigger than the surface pressure, the speed of sound in the
wall would be superluminal.  The conclusion is that the wall must have a
mass comparable to the magnitude of $\varepsilon$, or larger. This will make
$E_0$ positive.  Since an exact cancellation between wall
mass and vacuum energy is unlikely, we expect  $E_0$ to be of order
$|\bar\alpha|\hbar/R$, or larger. Thus, henceforth,
we simply write $E_0=\alpha N\hbar/R$ and
assume $\alpha>0$.

The formula for entropy, Eq.~(\ref{e22}) is now
\begin{equation}
S_{\rm rad}=
(4\surd 2/135)N^{1/4}[45\pi (RE/\hbar - \alpha N)]^{3/4}. \label{e23}
\end{equation}
Now the function $f(x)=(x-a)^{3/4}/x$ has a maximum value of ${1\over
4}3^{3/4}a^{-1/4}$.  We may, therefore, deduce from Eq.~(\ref{e23}) the
inequality
\begin{equation}
S_{\rm rad}\leq{\surd 2\,\pi^{3/4}\over(135\alpha)^{1/4}}{RE\over \hbar}.
\label{e24}
\end{equation}
This bound for blackbody radiation in a box is of the same form as bound
(\ref{e1}). Indeed, since typically $\alpha\sim 10^{-3}-10^{-2}$
for a single field \cite{Luk,Unwin},
the numerical coefficient here is $\sim 3-5.5$ in harmony with bound
(\ref{e1}).  Most important, the bound here obtained from statistical physics
is independent of $N$, just as (\ref{e1}) is.  The physical mechanism
 is that some of the
energy $E$ is inert energy whose magnitude depends on $N$:
$E>E_0=\alpha N\hbar/R$.  Thus the factor $RE/\hbar$ actually grows with $N$.
But the point is that when we state  an entropy bound in terms of $R$ and
$E$,  we do not have to worry about how big $N$ is; the bound will
take care of that automatically.

At this junction we return to the issue of the allowed range of $\eta$.
We have already seen that when $N$ is a few, and the box is standardly
oriented,
$\eta\ll 1$. Let us now analyze the case when  $N$ is large
for a spherical box of radius $R$.  Then we may use our result that
$E>E_0=\alpha N\hbar/R$ in the definition (\ref{e12}) by taking $A=4\pi R^2$
and $b=2R$.  We get
\begin{equation}
\eta^3={N\hbar\over 1440\pi ER}<{1\over 1440\pi\alpha},  \label{e25}
\end{equation}
which shows that $\eta$ does not grow indefinitely with $N$. Furthermore,
unless $\alpha$ is much smaller than the typical value for separate fields,
$\sim 10^{-2}-10^{-3}$, $\eta$ is indeed
 small compared to unity, thus allowing the
arguments following Eq.~(\ref{e12}) to yield the entropy bound from an
appeal to black hole physics. And this bound is, of course, consistent
with Eq.~(\ref{e24}).

The case of a nonspherical box is much harder to analyze because  $E_0$
depends not only on the typical dimension of the box, but also on axis
ratios \cite{Luk}.  However, it seems plausible that $\eta$ will also be
small in those cases  provided the various box dimensions are not too
different.

\subsection{Irrelevance of Many Dimensions}

Not only does the proliferation of particle species increase the entropy for
given energy, proliferation of spatial dimensions has the same effect.
Evidently the more the dimensions, the more ways there are to split up the
energy, so that a higher entropy is obtained.  We might thus naively expect
a bound like (\ref{e1}) to be violated as the number of dimension increases
without bound.  As we now show, this conclusion would be premature.
Consider in $D$
flat spatial dimensions a spherical space  of radius $R$ into which we dump
energy $E$.  What can we say about the entropy $S(E)$ as $D$ grows ?
Evidently maximal $S(E)$ corresponds to the excitation of all existing field
species with like inverse temperature $\beta$.  The description of our fixed
energy system in terms of temperature (thermodynamic regime)
is tenable provided $E$ is large enough
that energy fluctuations in the canonical ensemble for $\beta$
are small.  In practice this means $\beta\hbar/R\ll 1$ (many wavelenghts small
compared to $R$ are thermally excited), which we assume to be true. We shall
simplify matters by ignoring massive species.  This corresponds to the case
that $\beta$ times any of the rest masses is large.

The volume of a sphere of radius $r$ in $D$ dimensions is \cite{Huang}
\begin{equation}
V_D(r)={2\pi^{D/2}r^D\over D\Gamma(D/2)},   \label{e26}
\end{equation}
where $\Gamma$ denotes the Euler gamma function.  Consequently, the volume in
frequency space of the shell $(\omega,\omega+d\omega)$ is
\begin{equation}
dV_D(\omega)=D[V_D(\omega)/\omega]d\omega. \label{e27}
\end{equation}
The mean thermal energy  in the sphere from one helicity degree of
freedom is
\begin{equation}
E=V_D(R)\int_0^\infty{{\hbar\omega\, dV_D(\omega) \over
(e^{\beta\hbar\omega}\mp 1)(2\pi)^D}},   \label{e28}
\end{equation}
where upper (lower) signs correspond to boson (fermion) fields.
Using Eqs.~(\ref{e26}-\ref{e27}),  and
\begin{equation}
\int_0^\infty{{x^D\,dx\over e^{x}\mp 1}}=\zeta(D+1)\Gamma(D+1)\times
\cases{1\ \ &
for bosons;\cr 1-2^{-D}\ \  &for fermions.\cr}  \label{e29}
\end{equation}
where $\zeta(z)$ is Riemann's zeta function, we can cast the mean energy of all
massless species in the form
\begin{equation}
E={N\zeta(D+1)\Gamma(D)R^D\over 2^{D-3}[\Gamma(D/2)]^2\beta^{D+1}\hbar^D},
\label{e30}
\end{equation}
where $N$ is the number of massless species (massless nonscalar bosons
contribute one to $N$, while massless fermions contribute
$2^{-1}-2^{-(D+1)}$).

Likewise, we can write the thermal entropy of one helicity degree of freedom as
\begin{equation}
S=V_D(R)\int_0^\infty{\left[\mp\ln(1\mp e^{-\beta\hbar\omega})
+{\beta\hbar\omega\over
e^{\beta\hbar\omega}\mp 1}\right]{dV_D(\omega)\over (2\pi)^D}}.
\label{e31}
\end{equation}
After integration by parts, the first term in square brackets is seen to
reduce, by virtue of Eq.~(\ref{e27}), to $D^{-1}$ times the second.
Comparing the result for the entropy of all helicity degrees of freedom with
Eq.~(\ref{e28}) we see that
\begin{equation}
S_N=(1+1/D)\beta E.  \label{e32}
\end{equation}
The special case for $D=3$ of this equation is well known.

As mentioned, the condition  $\beta\hbar/R\ll 1$ must be respected in
order for the above continuum treatment to make sense.  Solving
Eq.~(\ref{e30}) for  the relevant quantity we have
\begin{equation}
\beta\hbar/ R=C_D (N\hbar/RE)^{1\over D+1},
\label{e33}
\end{equation}
with
\begin{equation}
C_D\equiv \left[{\zeta(D+1)\Gamma(D)\over
2^{D-3}[\Gamma(D/2)]^2}\right]^{1\over D+1}. \label{e34}
\end{equation}
Numerically it is found that $C_D$ decreases with $D$:
$C_D=(1.34,1.29,1.25, \dots, 1)$ for $D=(1,2,3,\dots, \infty)$.  Hence the
condition on $\beta$ amounts to $RE/N\hbar\gg 1$.  So, just as in
Sec.~III.A,  we find here that $N$ constrains the value of $RE$.  Substituting
Eqs.~(\ref{e33}-\ref{e34}) into Eq.~(\ref{e32}) we get
\begin{equation}
S_N=C_D (1+1/D) N^{1\over D+1} (RE/\hbar)^{D\over D+1}. \label{e35}
\end{equation}
The case $D=3$ of this should be compared with Eq.~(\ref{e23}); the only real
difference is that here $E$ does not include vacuum energy.

Let us now take the limit $D\rightarrow \infty$ while keeping $N$ fixed. In
this limit
$\zeta(D+1)\rightarrow 1$ and $\Gamma(D+1)\rightarrow
\sqrt{2\pi}e^{-D\,}D^D$, so we find rigorously that
\begin{equation}
\lim_{D\rightarrow\infty}S_N=RE/\hbar.   \label{e36}
\end{equation}
Thus despite the many dimensions, and the arbitrary number of species $N$,
this entropy is consistent with bound (\ref{e1}).  As in Sec.~III.A, the
dependence on $N$ has dropped out except for the fact that our end result is
valid only if $E$ is large enough on a scale set by $N$.  And contrary to
intuition, in the thermodynamic limit the large phase space that opens up
because of the multiplicity of dimensions does not  help the entropy to surpass
the entropy bound (\ref{e1}).

What happens for $D$ finite ?  We rewrite Eq.~(\ref{e35}) as
\begin{equation}
S_N=C_D(1+1/D)(N\hbar/RE)^{1\over D+1}\,RE/\hbar.  \label{e37}
\end{equation}
Numerically $C_D(1+1/D)$ decreases monotonically from $2.89$ at $D=1$ to
unity as $D\rightarrow\infty$.  Because $N\hbar/RE\ll 1$ in the thermodynamic
regime, it is clear that the entropy $S_N$ conforms to bound (\ref{e1}) for any
$D$.  We must stress that this result, like the previous one, is not
guaranteed if the condition for the thermodynamic regime fails.  In that case
one cannot rely on continuum formulae like Eqs.(\ref{e31}-\ref{e32}),
and must take recourse to numerical calculations of the energy
distribution
of quantum states in a cavity.  As mentioned, for $D=3$ such calculations
have been made \cite{Numer}, and fully confirm bound (\ref{e1}).  When
generalizing these to arbitrary $D$ and $N$, care should be taken to
include vacuum energies.

\section{INFORMATION CAPACITY OF REMNANTS}

Remnants of black holes have been suggested as a resolution of the black hole
information paradox (for reviews see Ref. \cite{info}).  In one version of this
idea \cite{remnants}, the black hole stops evaporating when its dimension
approaches the  Planck scale.  It is hypothesized that this remnant quiescent
object retains the large information that went down the initial black hole upon
its formation.  With a notable exception \cite{Gidd}, the arguments for and
against Planck scale remnants  have not confronted {\it quantitatively\/} the
question of how much information can really be contained in a space which looks
so small from the outside, and has a correspondingly tiny externally measured
mass energy.

Now, if a system contains information $I$ (natural units), then it must
have at its disposal at least $n=e^I$ internal states.  When looked at
in coarse grainning, such a system
will be ascribed entropy $S={\rm max}(\ln n)\geq I$.
Now, the entropy bound (\ref{e1}) must also bound the
entropy of  a gravitating system since we never specified in Sec.~II
details about the box's interior.  Applied to a remnant the bound would
predict that for radius $\ell_{\rm Planck}$ and mass
$\hbar/\ell_{\rm Planck}$ its entropy is $2\pi$ at most.  The remnants
information content is thus just a few bits.
This makes Planck--size remnants irrelevant for the information
paradox (the system that collapsed to the black hole which fathered the
remnant may have been specified by trillions of bits).

Giddings introduced the idea of large remnants \cite{Gidd} to overcome
problems like this.  The idea is that
each black hole decays to a remnant of a different size, which is deemed
capable of retaining the initial information.  Now, for a Schwarzschild
black hole
of initial mass $M_0$, the initial black hole entropy $S_{\rm bh}=
4\pi M_0^2$ measures the total information that becomes hidden at the moment of
collapse.  As argued above, Eq.~(\ref{e1}) should bound the information that
can be held by the remnant whose mass and radius are $E$ and $R$.  We assume
$R\sim 2E$ because the remnant must be a strongly gravitating object.  Thus
for the remnant to succesfully retain the initial information, we need $E\geq
M_0$ which is impossible since some evaporation must have taken place.  The
point is that black holes saturate the entropy (information) bound
so that the equivalent amount of information cannot be held by a lighter
strongly gravitating object.

Clearly the role of remnants, Planck size or larger, might be rescued if they
were not to respect the entropy bound (\ref{e1}).  We have argued that the
bound must apply to any object that can be lowered to the horizon of a black
hole.  Perhaps a remnant, by virtue of its black hole nature, cannot be
supported
in the requisite way to be lowered.  However that may be, any remnant can still
be dropped freely into a bigger extreme RN black hole of mass $M$.
Repeating the
argument at the beginning of  Sec.~III.A,
and arguing that the  information--coding states of the
remnant contribute towards coarse grained entropy,  we see that
the procedure bounds the  entropy, and hence the information capacity, of the
remnant by $\pi (2M)^{3/2}E^{1/2}$  (otherwise the GSL gets violated in
the remnant black hole merger).

Now $M$ has to be bigger than $E$ so that the remnant can be absorbed, but
it does not have to be arbitrarily large.  Thus this independent argument
bounds the remnant's information capacity by $\gamma E^2$ with $\gamma$ perhaps
a few times $2^{3/2}\pi$.  If the remnant has descended from a neutral black
hole of initial mass $M_0$, it must retain information $16\pi M_0^2$ to
do its job as information repository in the resolution of the
information paradox.  This information
will conform to the bound we have just set provided
$E>4(\pi/\gamma)^{1/2}M_0$.  We see that the remnant cannot be much
lighter
than $M_0$.  However, there seems to be no reason for black hole evaporation
to turn off as soon as the black hole has lost a moderate fraction of its
mass.  This problem dramatizes the difficulty the  remnant scenario has in
resolving the black hole information paradox.

The usual argument for remnants as information repositories is that they can
retain large information despite their small dimension because they contain a
very large internal space in the form of a throat or horn.  According to this
argument, a bound like (\ref{e1}), which is stated in terms of external
dimensions alone, cannot apply.  But we have just seen that if a remnant
had an information capacity well above that indicated by bound (\ref{e1}) in
terms of its external dimension, it would cause a violation of
the GSL, were it dropped into a large black hole. This point might be
countered if the infall of the remnant produces a singularity, thus making the
GSL moot.  However, in that case the whole role of remnants as stable
information repositories is put in doubt.  The conclusion must be that black
remnants, as discussed heretofore, cannot resolve the information paradox.

\acknowledgments

I thank Ted Jacobson, Don Page and Bob Wald
 for a number of remarks, Leonard Parker for
suggestions, Jim Hartle for his interest in the subject,  Hartle
and Gary Horowitz for hospitality in Santa Barbara, and the Aspen Center
for Physics where the draft of this paper was completed.

\end{document}